\documentclass[3p,preview]{elsarticle}

\journal{Scripta Materialia}

\usepackage{amsmath}
\usepackage{xcolor}
\usepackage{mathtools}
\usepackage{soul}

\definecolor{darkgreen}{rgb}{0.1,0.8,0.1}

\newcommand{\fig}{Figure}

\newcommand{\sub}[1]{\ensuremath{_{\textrm{#1}}}}

\begin{document}

\begin{frontmatter}

\title{Demonstrating the potential of Accurate Absolute Cross-grain Stress and Orientation correlation using Electron Backscatter Diffraction}

%
%

\author[mymainaddress]{Tijmen Vermeij}
\address[mymainaddress]{Dept. of Mechanical Engineering, Eindhoven University of Technology, 5600MB Eindhoven, The Netherlands}
\author[secondaddress]{Marc De Graef}
\address[secondaddress]{Dept. of Materials Science and Engineering, Carnegie Mellon University, Pittsburgh PA 15213-3890, USA}
\author[mymainaddress]{Johan Hoefnagels*}
\cortext[mycorrespondingauthor]{Corresponding author}
\ead{j.p.m.hoefnagels@tue.nl}

\begin{abstract}
We report a first exploration of High-angular-Resolution Electron Backscatter Diffraction, without using simulated Electron Backscatter Diffraction patterns as a reference, for absolute stress and orientation measurements in polycrystalline materials. By co-correlating the pattern center and fully exploiting crystal symmetry and plane-stress, simultaneous correlation of all overlapping regions of interest in multiple direct-electron-detector, energy-filtered Electron Backscatter Diffraction patterns is achieved. The potential for highly accurate measurement of absolute stress, crystal orientation and pattern center is demonstrated on a virtual polycrystalline case-study, showing errors respectively below $20$ MPa (or $10^{-4}$ in strain), $7\times10^{-5}$ rad and $0.06$ pixels. DOI: \url{https://doi.org/10.1016/j.scriptamat.2018.11.030}
\end{abstract}


\end{frontmatter}
\textbf{Graphical Abstract}
\begin{figure}[h]
	\centering
	\includegraphics[width=\textwidth]{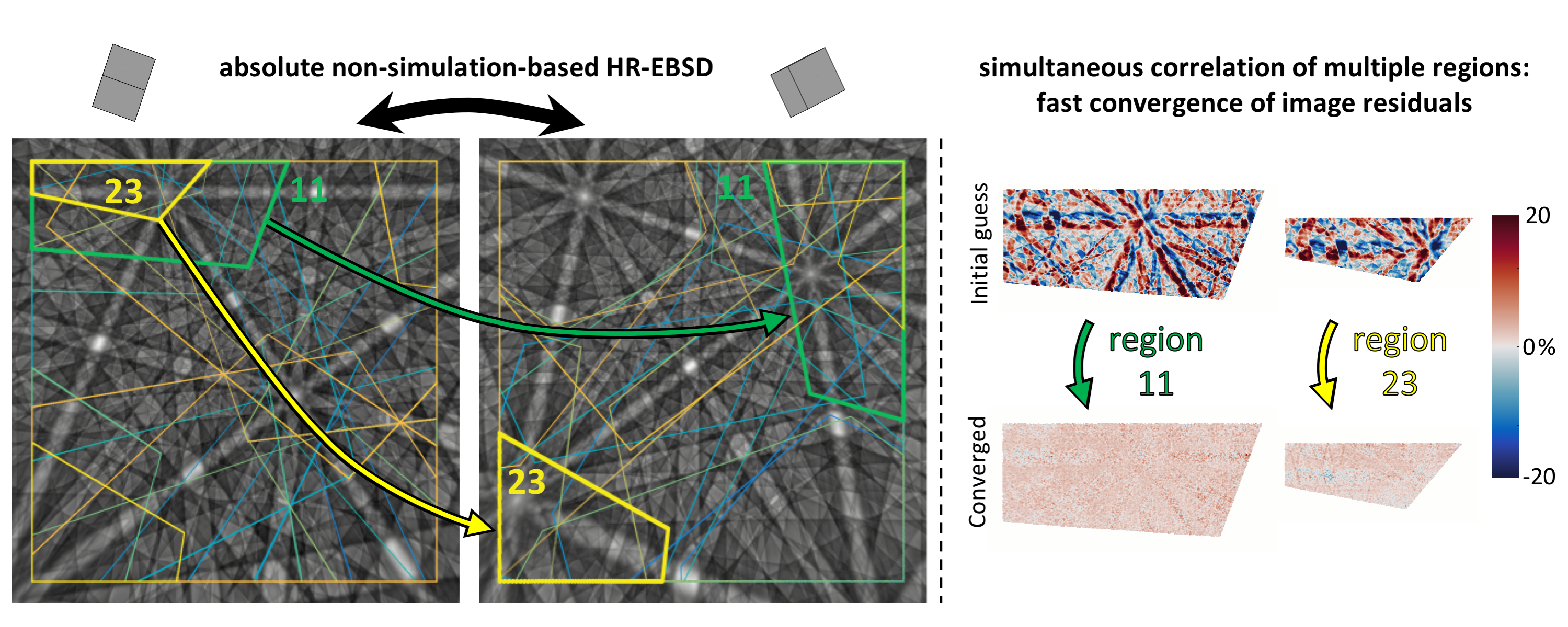}
\end{figure}

A novel and accessible technique that can provide unprecedented details of grain boundaries (GBs) in polycrystalline materials, particularly higher accuracy of GB misorientation and GB compatibility stresses and strains at high spatial resolution, may (i) provide fundamental understanding of GB deformation mechanisms, such as dislocation-GB interactions (pile-up, transmission, absorption, void nucleation, etc.) \cite{bachurin2010dislocation, MALYAR2017312}, twinning \cite{zhu2012deformation}, (nano-)grain rotations \cite{wang2014grain} and GB sliding \cite{bobylev2010cooperative}, and (ii) open up new pathways to design novel high-performance alloys \cite{HIRTH2011586} such as transformation- and twinning-induced plasticity steels \cite{miyamoto2009precise}, shape memory alloys \cite{wen2014large}, self-healing alloys \cite{xu2013healing, ulvestad2017self}, nano-laminated steels \cite{koyama2017bone}, metallic glasses \cite{greer2011plasticity}, metastable high-entropy alloys \cite{li2016metastable}, etc. \cite{frolov2013structural,zhang2017complexion}.

Although quantitative nano-scale crystallography \mbox{\cite{robinson2009coherent, hofmann2013x, sedmak2016grain, abdolvand2018strong}} is actively researched by synchrotron based 3D X-ray diffraction \mbox{\cite{Poulsen:nb5032}}, we propose a new variation on the more accessible High-angular-Resolution Electron Backscatter Diffraction (HR-EBSD) method. While automated 2D-Hough transform-based EBSD indexing is the standard for texture analysis \cite{adams1994}, HR-EBSD, pioneered by Wilkinson et al. \cite{wilkinson_high-resolution_2006}, provides an extension to simultaneously measure the stress state by subset-based Digital Image Correlation (DIC) of the Electron Backscatter Patterns (EBSPs) to a reference EBSP. In \textit{absolute} HR-EBSD, a \textit{simulated} EBSP is used as reference \cite{kacher_braggs_2009, fullwood2015validation, jackson2016performance}, yet, these methods suffer from uncertainties in the calibration of the experimental geometry, specifically the Pattern Center (PC) location \cite{maurice_comments_2010, kacher2010reply, britton_factors_2010, mingard_towards_2011, maurice_method_2011, basinger2011pattern, alkorta_limits_2013, britton_assessing_2013}, and inaccurate simulation of experimental EBSP features \cite{PhysRevB.97.134104}, although developments are ongoing \cite{alkorta_improved_2017}. In contrast, \textit{relative} HR-EBSD is much more accurate with errors in elastic strains of ${\sim}10^{-4}$ \cite{villert_accuracy_2009, plancher_accuracy_2016}; however, this approach requires one EBSP in each grain as reference, thus only yielding stress \textit{gradients} inside grains, with maximum misorientations of ${\sim}10^{\circ}$ \cite{maurice_solving_2012, britton_high_2012}. As typically the full stress state is not known anywhere in a grain, \textit{absolute} stress level determination at all points is impossible, let alone correlation across GBs.

\begin{figure}
 \centering
	\includegraphics[width=12cm]{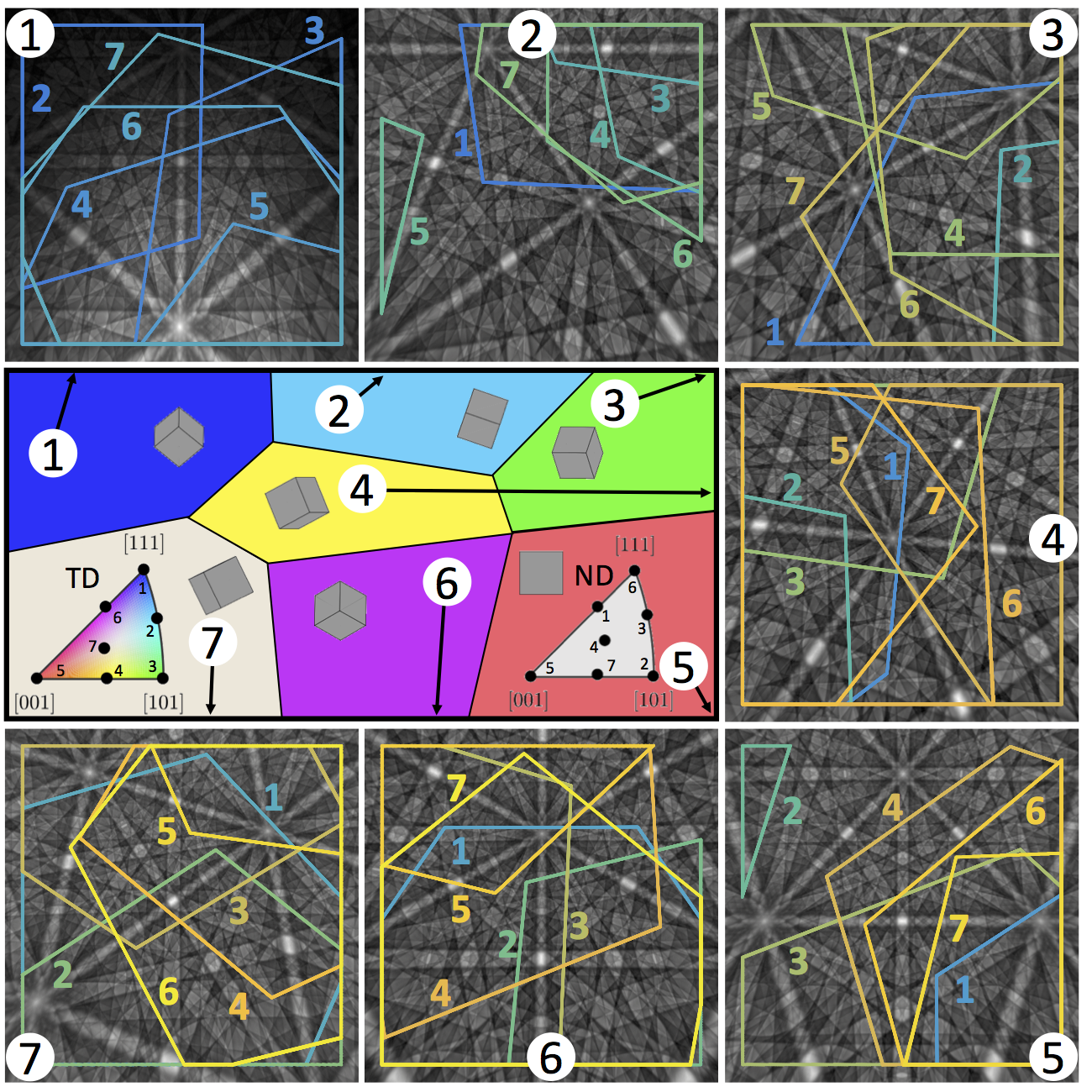}
	\caption{Case-study of an artificial polycrystalline microstructure of BCC Ferrite with large grain misorientations (see rotated cubes). The 7 grains fully cover the transverse direction (TD) and normal direction (ND) inverse pole figure (IPF), and are selected as far apart as possible to ensure that (cubic symmetric) polycrystalline microstructures encountered in practice will not show larger misorientations than the grains tested here. Their EBSPs, generated by dynamical simulations and stressed (elastically strained) according to table \ref{tab:table1}, are corrected by division of an average background, collected over many grain orientations (similar to experiments). 21 gROIs are drawn as colored lines in the EBSPs, each illustrating an overlap between a pair of EBSPs. The gROI label numbers denote the paired EBSPs. Note that EBSP\sub{$1$} is shown with the original background.}
	\label{fig:fig1}
\end{figure}

\begin{figure}
 \centering
	\includegraphics[width=9cm]{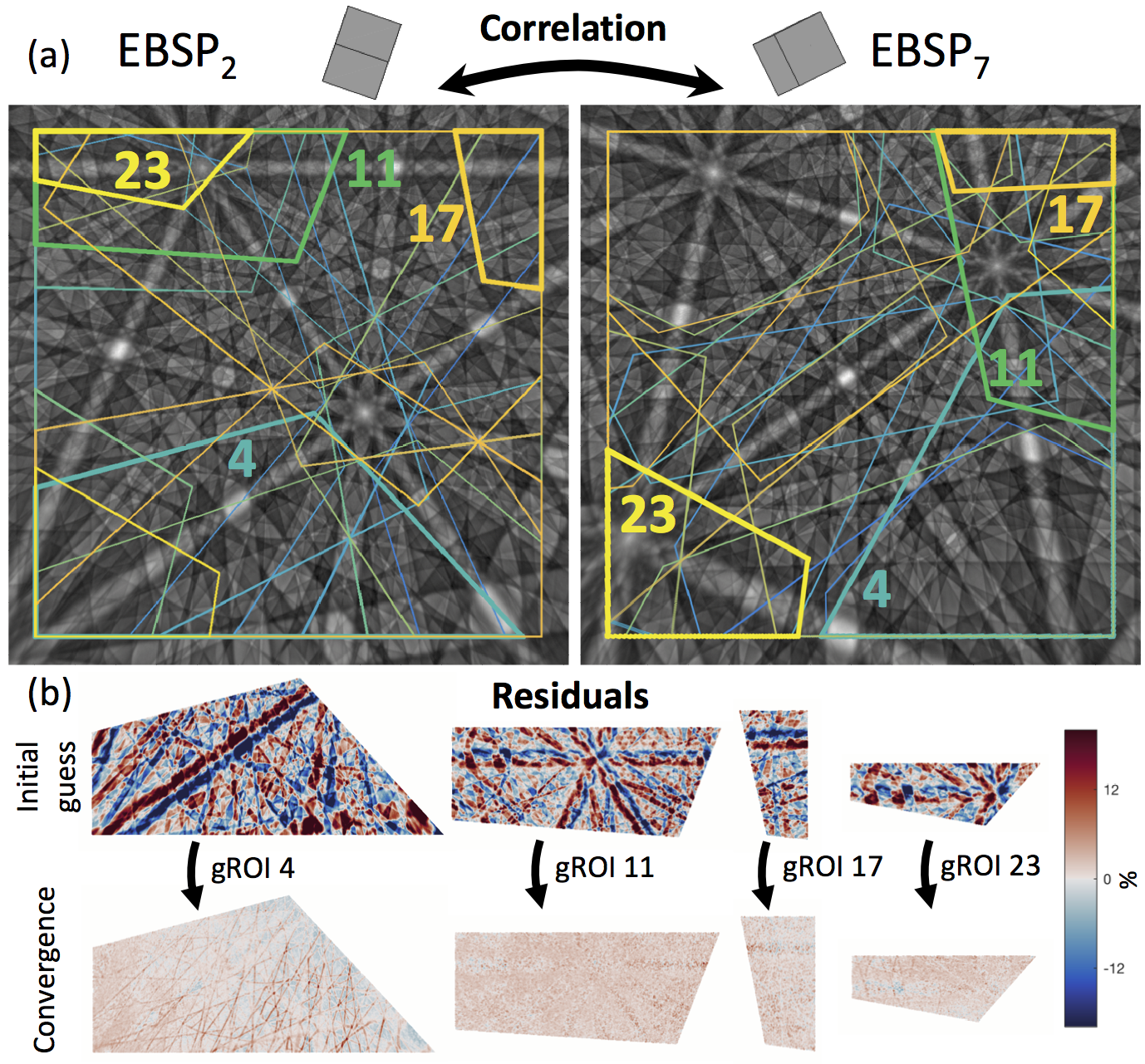}
	\caption{Full cubic symmetry assisted correlation of EBSP\sub{$2$} with EBSP\sub{$7$}, concurrent to \fig~\protect\ref{fig:fig3}c. (a) All 23 non-zero gROIs in both EBSPs, labeled by color. (b) Examples of EBSP residual fields ($r_{2,7,4}$, $r_{2,7,11}$, $r_{2,7,17}$, $r_{2,7,23}$), highlighted in (a), at initial guess and after convergence.}
	\label{fig:fig2}
\end{figure}

This calls for a paradigm shift in how \textit{absolute} HR-EBSD is approached. First, for a polycrystalline structure, all the global Regions Of Interest (gROIs), i.e., overlapping areas, between each EBSP from each grain can be correlated at once to boost the sensitivity, as shown in \fig~\ref{fig:fig1} for the simple example of only 1 EBSP in each of 7 grains, constituting 21 EBSP pairs. Second, the sensitivity can be further enhanced by fully exploiting crystal symmetry, yielding up to 24 gROIs for \textit{each} EBSP pair (in the case of cubic symmetry), as shown in \fig~\ref{fig:fig2}a, thus resulting in a maximum total of 504 gROIs for the example of \fig~\ref{fig:fig1}, that can simultaneously be correlated. Hence, we report the first exploration of \textit{absolute} HR-EBSD to enable highly accurate identification of the \textit{absolute} stress tensor, crystal orientations and PC coordinates across GBs, \textit{without} using simulated EBSPs as reference. This is achieved by fully exploiting the recently proposed integrated DIC (IDIC) based HR-EBSD framework of Vermeij \& Hoefnagels \cite{VERMEIJ201844}, based on a consistent full-field one-step optimization approach instead of standard two-step subset-based HR-EBSD algorithms, while taking full advantage of the crystal symmetry, plane stress conditions and correlation of multiple gROIs. Thereby, full cross-grain correlations are explored and validated on a challenging virtual stressed polycrystalline case-study.

The determination of the correct set of Degrees of Freedom (DOFs), $\{ \lambda \}$, containing the stress and orientation per EBSP and the PC coordinates, is achieved by minimization of the brightness residual, $r_{i,j,s}$,
\begin{equation}
r_{i,j,s}\big(\vec{x}_i,\{ \lambda \}\big) = g_i\big(\vec{x}_i\big) - g_j\big(\vec{x}_i+\vec{u}_{i,j,s}(\vec{x}_i,\{ \lambda \})\big),
\end{equation}
for each gROI ($\Omega_{i,j,s}$) between each pair of EBSPs $g_i$ and $g_j$ subjected to the symmetry operator $s$ and defined by a displacement field $\vec{u}_{i,j,s}$ at pixel position $\vec{x}_i$ \cite{VERMEIJ201844, NME:NME2908,neggers_gradients, neggers_time-resolved_2015, ruybalid_comparison_2016}.
This multiple-gROI, multiple EBSP minimization yields:
\begin{equation}
\{ \lambda \} = \underset{\lambda}{\operatorname{argmin}} \ \sum\limits_{i=1}^{N-1}\ \sum\limits_{j=i+1}^{N}\ \sum\limits_{s=1}^{N_s} \ \int_{\Omega_{i,j,s}}[r_{i,j,s}(\vec{x}_i,\{ \lambda \})]^2\ \mathrm{d}\vec{x},
\end{equation}
where $\underset{\lambda}{\operatorname{argmin}}$ denotes the minimization with respect to the DOFs $\{ \lambda \}$, $N$ is the number of EBSPs in the correlation and $N_s$ is the number of different symmetry operators. The initial guess for $\{\lambda\}$ is iteratively updated during the optimization until convergence is met. Note that no EBSP is treated as an "undeformed" pattern; instead, the deformed EBSPs are correlated by considering their \textit{relative} deformation and orientation, which can be directly related to their \textit{absolute} deformation and orientation. Since EBSPs originate from a ${\sim}10$ nm thick volume directly beneath the traction-free specimen surface \cite{hardin_analysis_2015}, plane-stress is assumed, as is common in HR-EBSD literature. Generally, however, only the out-of-plane normal Cauchy stress component is constrained to zero ($\sigma_i^{33}=0$). In this work, however, following \cite{alkorta_improved_2017}, also the out-of-plane shear stress components are constrained, i.e. $\sigma_i^{13}=\sigma_i^{23}=0$, to maximize sensitivity for all $\{ \lambda \}$, while aiming to accurately measure the remaining in-plane stress components $\sigma_i^{11}$, $\sigma_i^{22}$, and $\sigma_i^{12}$. Additionally, for each EBSP\sub{$i$}, the crystal orientation is included in the DOFs as a set of three Euler angles, fully describing a rotation tensor $\mathbf{R}_i$ in the global specimen coordinate system. Furthermore, the DOFs of one set of global (or \textit{absolute}) PC coordinates (i.e., location $\vec{x}^{pc}$ from the top-left in the EBSP and detector distance, $dd$, both defined in pixels or px) is added to the optimization routine, while the \textit{relative} PC changes between EBSPs are assumed to be known from the beam shifts. Altogether, the list of DOFs consists of:
\begin{equation}
\{ \lambda \} = \{...,\sigma^{11}_i,\sigma^{22}_i,\sigma^{12}_i,Eu_i^X,Eu_i^Y,Eu_i^Z,...,pc^x,pc^y,dd\},
\end{equation}
with $1\leq i\leq N$.

Next, we need the displacement field $\vec{u}_{i,j,s}(\vec{x}_i,\{ \lambda \})$ for each gROI to perform the correlation. Let us consider EBSP\sub{$i$}, consisting of a field of gray values $g_i$, originating from a cubic symmetric material point $i$ which has a certain crystal orientation, defined by rotation tensor $\mathbf{R}_i$, and is stressed by Cauchy stress tensor $\mathbf{\sigma}_i$, both defined in the global specimen coordinate system. When comparing any two EBSPs in a (poly)crystalline microstructure, e.g., EBSP\sub{$i$} and EBSP\sub{$j$}, a pixel in EBSP\sub{$i$} with position $\vec{x}_i$, within gROI $\Omega_{i,j,s}$, can be found in EBSP\sub{$j$} at position $\vec{x}_j=\vec{x}_i+\vec{u}_{i,j,s}(\vec{x}_i,\{ \lambda \})$. As a typical example, \fig~\ref{fig:fig1} shows dynamically simulated EBSPs for each grain, in which the overlapping areas, or gROIs, are automatically calculated for each pair of EBSPs, based on the displacement field $\vec{u}_{i,j,s}(\vec{x}_i,\{ \lambda \})$, which was derived in \cite{VERMEIJ201844} as function of the DOFs $\{ \lambda \}$ and is based on the EBSP formation geometry:
\begin{equation}
\vec{u}_{i,j,s}(\vec{x}_i,\{ \lambda \})=\frac{dd_j}{\vec{e}_z\cdot \mathbf{F_{r}}\cdot\vec{x}_i^{''}} \Big( \mathbf{F_{r}}\cdot\vec{x}_i^{''}-\big( \vec{e}_z\cdot\mathbf{F_{r}}\cdot\vec{x}_i^{''}\big) \vec{e}_z \Big) +\vec{x}_j^{pc} - \vec{x}_i,
\label{eq:eq1}
\end{equation}
wherein we define $\vec{x}_i^{''}=dd_i \vec{e}_z+\vec{x}_i-\vec{x}_i^{pc}$, with $\vec{e}_z$ a normal unit vector on the detector screen. The relative deformation gradient tensor equals $\mathbf{F_{r}}=\mathbf{F_{t}}^T\cdot \mathbf{F}_{i,j,s}\cdot \mathbf{F_{t}}$, in which $\mathbf{F_{t}}$ is the rotation tensor specifying the specimen tilt, while $\mathbf{F}_{i,j,s}=\mathbf{F}_j\cdot\mathbf{R}_s\cdot \mathbf{F}_i^{-1}$ denotes the \textit{relative} deformation gradient tensor between material point $i$ and $j$. $\mathbf{F}_i$ and $\mathbf{F}_j$ are the \textit{absolute} deformation gradient tensors of material points $i$ and $j$, with respect to an undeformed crystal that is aligned with the specimen coordinate system, which are uniquely defined by crystal orientation $\mathbf{R}_i$ and $\mathbf{R}_j$ and right stretch tensors, $\mathbf{U}_i$ and $\mathbf{U}_j$, as e.g. $\mathbf{F}_i=\mathbf{R}_i\cdot \mathbf{U}_i$. The additional rotation tensor $\mathbf{R}_s$ is one of a number of possible symmetry rotation operators specific to the symmetry of the crystal system. $\mathbf{R}_s$ can thus vary to result in a number of possibilities for $\mathbf{F_{r}}$, resulting in the existence of multiple gROIs between a set of EBSPs, as demonstrated in \fig~\ref{fig:fig2}a. This feature has so far never been exploited in HR-EBSD. Finally, we relate the stress state of the crystal in its current configuration, i.e. the Cauchy stress tensor $\mathbf{\sigma}_i$, to $\mathbf{R}_i$, $\mathbf{U}_i$ and the fourth order elastic stiffness tensor $\prescript{4}{}{\textbf{C}}$ \cite{VERMEIJ201844}:
\begin{equation}
\mathbf{\sigma}_i=\frac{\mathbf{R}_i\cdot\mathbf{U}_i}{\mathrm{det}(\mathbf{R}_i\cdot\mathbf{U}_i)}\cdot\prescript{4}{}{\textbf{C}}:\frac{1}{2}\big((\mathbf{R}_i\cdot\mathbf{U}_i)^T\cdot \mathbf{R}_i\cdot\mathbf{U}_i-\textbf{I}\big)\cdot\big(\mathbf{R}_i\cdot\mathbf{U}_i\big)^T.
\label{eq:eq2}
\end{equation}
This non-linear equation is solved iteratively for $\mathbf{U}_i$.

The performance in terms of flexibility, robustness and accuracy of the novel non-simulation-based \textit{absolute} HR-EBSD framework is evaluated on a challenging case-study of a virtual stressed polycrystalline microstructure, explained in \fig~\ref{fig:fig1} and table~\ref{tab:table1}. The 12 bit EBSPs of $1000\times1000$ px, with realistic background profiles, are dynamical simulated for a $20$ keV incident electron beam, using EMsoft \cite{callahan_de_graef_2013, singh2017emsoft}, based on a Monte Carlo simulation of the electron depth, energy, and intensity profile variation. Using appropriate lattice parameters, corresponding to the elastically strained (i.e., stressed) unit cell for the required crystal orientation, each EBSP is generated for a direct electron EBSD detector \cite{wilkinson2013direct} with $19.5$ keV energy thresholding \cite{vespucci2015digital}, Gaussian noise level of $2\%$, and PC coordinates of $pc^x{\approx}500$ px, $pc^y{\approx}300$ px and $dd{\approx}500$ px, with variations to simulate electron beam scanning.

 \begin{table}
 \caption{\label{tab:table1}Applied Von Mises (VM) stress and randomly varied in-plane components in GPa for each EBSP, while the out-of-plane stress components are constrained to zero. Note that EBSP\sub{$1a$} has the same crystal orientation as EBSP\sub{$1$}.}
 \centering
 \begin{tabular}{ l| r r r r}

  &$\sigma^{VM}$&$\sigma^{11}$&$\sigma^{22}$&$\sigma^{12}$\\ \hline
EBSP\sub{$1$}& 1& 0.435    &-0.181	&0.482 \\
EBSP\sub{$1a$}&0.5& -0.453	&-0.443	&-0.129\\
EBSP\sub{$2$}& 1&-0.129	&0.920	&-0.007 \\
EBSP\sub{$3$}& 1&0.147	&-0.478	&-0.476 \\
EBSP\sub{$4$}& 1&-0.221	&0.667	&0.345 \\
EBSP\sub{$5$}& 1&-0.888	&0.018	&0.255\\
EBSP\sub{$6$}& 1&-0.522	&-0.431	&0.505\\
EBSP\sub{$7$}& 1&0.710	&-0.368	&0.181 \\

 \end{tabular}
 \end{table}

\begin{figure}
 \centering
	\includegraphics[width=12cm]{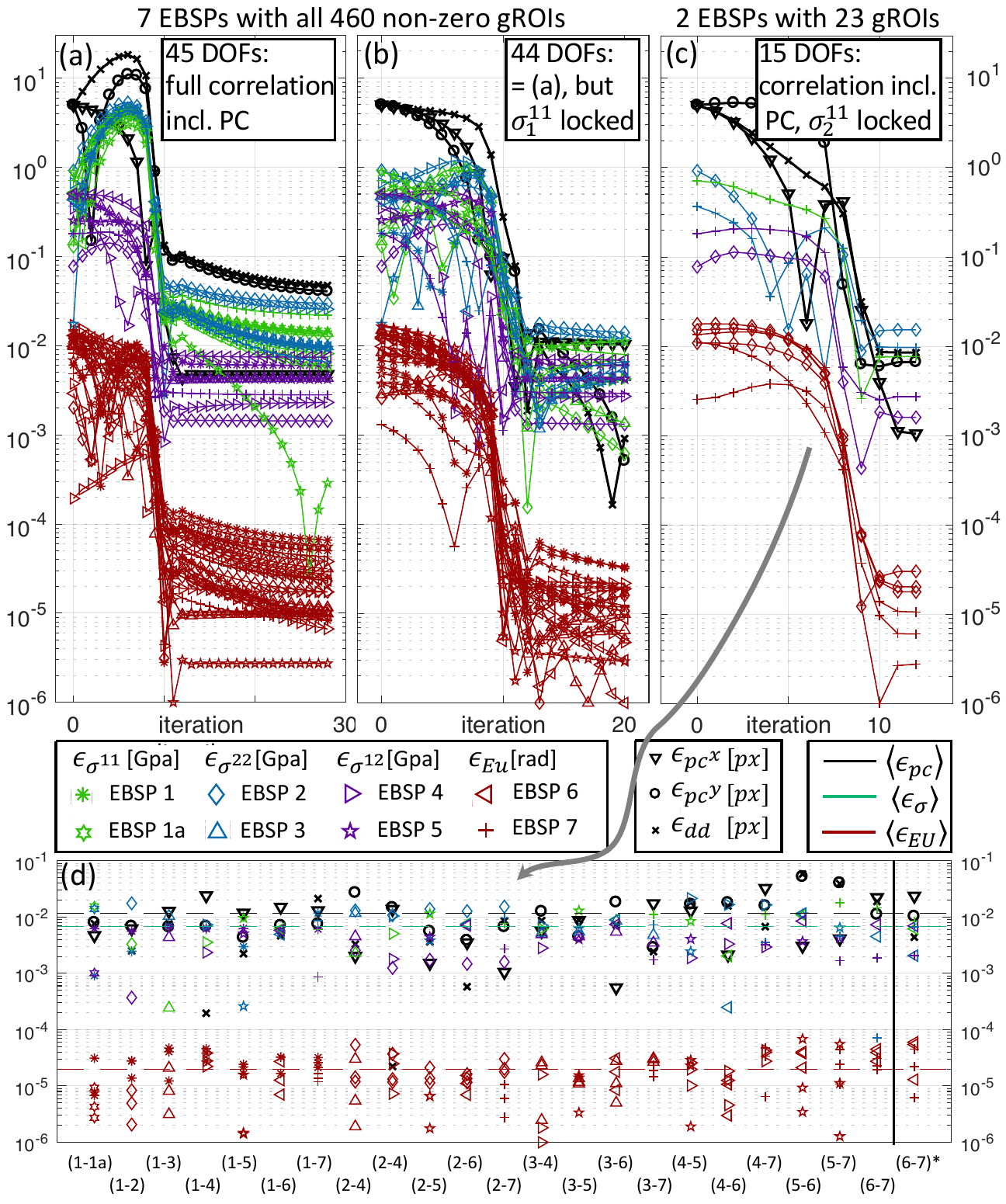}
	\caption{Performance evaluation of the cross-grain \textit{absolute} HR-EBSD algorithm. (a-b-c) Convergence behavior, in absolute errors, of simultaneous correlation of multiple EBSPs, including all 3 orientation and in-plane stress DOFs per EBSP, as well as the global PC DOFs (a) of all 7 EBSPs, (b) of all 7 EBSPs, with $\sigma^{11}_1$ assumed known and (c) of EBSP\sub{$2$} and EBSP\sub{$7$} (corresponding to \fig~\protect\ref{fig:fig2}), with $\sigma^{11}_i=\sigma^{11}_2$ assumed known. (d) Converged absolute errors, for correlation (similar to (c), with $\sigma^{11}_i$ known) of all combinations (i-j) of 2 EBSPs, with dashed lines showing the mean absolute errors. For combination (6-7)*, only $\sigma^{11}_i/\sigma^{11}_j=\sigma^{11}_6/\sigma^{11}_7$ is assumed known.}
	\label{fig:fig3}
\end{figure}

The accuracy of the non-simulation-based \textit{absolute} HR-EBSD algorithm is quantified by the absolute error metric $\epsilon_{\alpha} = |\alpha-\alpha^{\mathrm{ref}}|$, where $\alpha$ is a DOF and $\alpha^{\mathrm{ref}}$ the simulated reference value. The absolute errors of the stresses, orientations and PC coordinates are, respectively, expressed in units of GPa, radians and pixels (px) in \fig~\ref{fig:fig3}. To test robustness against experimental uncertainties, all virtual tests are initialized with a large offset in DOFs: a random orientation error of $1^{\circ}$, zero stress and $5$ px PC errors. A full correlation of the $7$ EBSPs, using all 460 available (out of maximum 504) gROIs in a single optimization step, with all orientation and in-plane stress components of the $7$ EBSPs and the $3$ global PC coordinates for a total of $45$ DOFs, results in convergence with low maximum errors in stress, orientation and PC of $30$ MPa, $10^{-4}$ rad and $0.1$ px, respectively, see \fig~\ref{fig:fig3}a. Complete correlation of all DOFs has not been achieved in the literature, yet, extensive testing showed that this is only possible when at least $5$ highly misoriented EBSPs are included in the correlation. This demonstrates the importance of the here-proposed paradigm shift to simultaneously correlate many gROIs from multiple EBSPs, in our flexible IDIC formulation, which would be unfeasible for the conventional two-step subset-based HR-EBSD algorithms.

When even higher accuracy is desired, a small assumption on the in-plane stress state can be included. Often one in-plane stress component in one EBSP is known due to stress relaxation at the specimen edge or by slit milling \cite{VERMEIJ201835}, or by attaining other insights on the stress state. This knowledge is sufficient to accurately correlate any combination of 2 (or more) EBSPs. This is demonstrated here by assuming knowledge of $\sigma^{11}$ for the first EBSP in each correlation, with \fig~\ref{fig:fig3}b and c, respectively, showing such a correlation for $7$ and only $2$ EBSPs, yielding higher accuracies in stresses, orientations and PC coordinates. Notably, for $7$ EBSPs, PC accuracies drop below $0.001$ px for $pc^y$ and $dd$, suggesting a highly stable correlation. \fig~\ref{fig:fig3}c shows the correlation between EBSP\sub{$2$} and EBSP\sub{$7$}, with the residual fields for 4 of the 23 gROIs shown in \fig~\ref{fig:fig2}b, demonstrating efficient minimization of the residual fields and optimization of the DOFs towards convergence. Subsequently, \fig~\ref{fig:fig3}d shows the accuracies of all available combinations of $2$ EBSPs. The successful correlation of EBSP\sub{$1$} with EBSP\sub{$1a$}, with same orientation yet different stress state, demonstrates that a misorientation between 2 EBSPs is not required.

Alternatively, stress components from different grains can be interlinked in the correlation by benefiting from, e.g., stress compatibility close to two sides of the GB. This approach is briefly tested in combination (6-7)* in \fig~\ref{fig:fig3}d, by assuming that the stress components $\sigma^{11}_6$ and $\sigma^{11}_7$ are linearly related, which is found to be equally accurate. Overall, errors of stress, orientation and PC components, respectively, remain below ${\sim}20$ MPa (or $<10^{-4}$ in strain), ${\sim}7\times10^{-5}$ rad and ${\sim}0.06$ px, while averaging ${\sim}7$ MPa, ${\sim}2\times10^{-5}$ rad and ${\sim}0.01$ px in this virtual case-study. Preliminary tests show equivalent accuracies when initial guesses vary or when 8 bit or 20\% noise EBSPs are correlated, while increasing $dd$ to $0.6$ decreases the accuracy by a factor of ${\sim}2$ (based on a preliminary test). No clear trend over the different combinations of EBSPs is observed, suggesting the powerful capability to correlate any pair of EBSPs under one limited assumption.

The accuracies achieved in this virtual case-study of dynamically simulated EBSPs, demonstrates that the IDIC based HR-EBSD method has the potential to perform \textit{absolute} HR-EBSD without using simulated EBSPs as reference, i.e., non-simulation-based. Direct comparison to state-of-the-art simulation-based \textit{absolute} HR-EBSD methods is currently not possible, as only experimental investigations are available in the literature without virtual or direct validation of accuracies. Importantly however, the level of accuracy of the \textit{relative} intergranular (cross-grain) strains and misorientations, also better than $10^{-4}$ in this work, has not been achieved, or even attempted, in (HR-)EBSD literature. Additionally, accurate measurement of PC coordinates, performed here alongside the correlation of stresses and orientations, is highly relevant and poses challenges to state-of-the-art \textit{absolute} \cite{britton_factors_2010, maurice_method_2011, basinger2011pattern, alkorta_limits_2013, alkorta_improved_2017} and even \textit{relative} \cite{britton_high_2012,VERMEIJ201844} HR-EBSD. Yet, experimental validation is required, preferably using energy-filtered direct electron EBSD detectors \cite{vespucci2015digital}, to study the effects of incident voltage, pattern background, detector noise, non-uniform gain, (relative) pattern quality, band anisotropy, uncertainties in the elastic constants, etc. Conventional EBSD detectors yield energy (and thus Kikuchi bandwidth) variations over the detector screen \cite{PhysRevB.97.134104} and can have problematic optical distortions \cite{mingard_towards_2011}, diminishing the method's practical accuracy. However, the flexible and consistent IDIC framework can be adapted to correct for the optical distortion by introducing hierarchical mapping functions that describe the interaction of the imaging process with the EBSP formation \mbox{\cite{MARAGHECHI2018144,maraghechi2018}}, allowing much room for further optimization. Finally, uncertainties in the specimen tilt cause errors in the absolute crystal orientation \mbox{\cite{nolze2007image}}, plane stress assumptions and relative PC coordinates \cite{plancher_accuracy_2016} though the relative PC error is negligible when scanning around a grain boundary (e.g. for a $10\times10 \ \mu m$ scan the error in $pc^y$ is ${\sim}0.0004$ px and ${\sim}0.002$ px, respectively, for tilt uncertainties of ${\sim}0.1^{\circ}$ and ${\sim}1^{\circ}$   ). Hence, it seems wise to include the specimen tilt as a DOF in the correlations or to explore other routes for accurate tilt calibration \mbox{\cite{nolze2007image}}.

In summary, we propose a non-simulation-based \textit{absolute} High-angular-Resolution EBSD approach that takes full advantage of plane stress assumptions, the crystal symmetry in an EBSD pattern, and the ability to correlate multiple regions of interest from multiple patterns in one optimization step. Validation on a challenging case-study of a virtual stressed polycrystalline, cubic-symmetric, microstructure shows, in theory, the potential to robustly and highly accurately provide the \textit{absolute} stress state and crystal orientation in all grains, while simultaneously determining the Pattern Center coordinates. Warranting further development and experimental validation, this method could open up possibilities of advanced high-resolution characterization of \textit{absolute} stress fields and \textit{absolute} orientations on both sides of grain boundaries in polycrystalline materials.

The authors thank Clemens Verhoosel, Hans van Dommelen and Marc Geers for discussions. MDG acknowledges financial support from an ONR Vannevar Bush Faculty Fellowship (N00014-16-1-2821).


\end{document}